\documentclass[12pt]{article}
\usepackage{amsfonts,amssymb,amsmath}
\usepackage[dvips]{epsfig}
\begin{document}

\def\ds{\displaystyle}
\def\beq{\begin{equation}}
\def\eeq{\end{equation}}
\def\bea{\begin{eqnarray}}
\def\eea{\end{eqnarray}}
\def\beeq{\begin{eqnarray}}
\def\eeeq{\end{eqnarray}}
\def\ve{\vert}
\def\vel{\left|}
\def\bpll{B\rar\pi \ell^+ \ell^-}
\def\ver{\right|}
\def\nnb{\nonumber}
\def\ga{\left(}
\def\dr{\right)}
\def\aga{\left\{}
\def\adr{\right\}}
\def\lla{\left<}
\def\rra{\right>}
\def\rar{\rightarrow}
\def\nnb{\nonumber}
\def\la{\langle}
\def\ra{\rangle}
\def\ba{\begin{array}}
\def\ea{\end{array}}
\def\tr{\mbox{Tr}}
\def\ssp{{\Sigma^{*+}}}
\def\sso{{\Sigma^{*0}}}
\def\ssm{{\Sigma^{*-}}}
\def\xis0{{\Xi^{*0}}}
\def\xism{{\Xi^{*-}}}
\def\qs{\la \bar s s \ra}
\def\qu{\la \bar u u \ra}
\def\qd{\la \bar d d \ra}
\def\qq{\la \bar q q \ra}
\def\gGgG{\la g^2 G^2 \ra}
\def\q{\gamma_5 \not\!q}
\def\x{\gamma_5 \not\!x}
\def\g5{\gamma_5}
\def\sb{S_Q^{cf}}
\def\sd{S_d^{be}}
\def\su{S_u^{ad}}
\def\ss{S_s^{??}}
\def\ll{\Lambda}
\def\lb{\Lambda_b}
\def\sbp{{S}_Q^{'cf}}
\def\sdp{{S}_d^{'be}}
\def\sup{{S}_u^{'ad}}
\def\ssp{{S}_s^{'??}}
\def\sig{\sigma_{\mu \nu} \gamma_5 p^\mu q^\nu}
\def\fo{f_0(\frac{s_0}{M^2})}
\def\ffi{f_1(\frac{s_0}{M^2})}
\def\fii{f_2(\frac{s_0}{M^2})}
\def\O{{\cal O}}
\def\sl{{\Sigma^0 \Lambda}}
\def\es{\!\!\! &=& \!\!\!}
\def\ar{&+& \!\!\!}
\def\ek{&-& \!\!\!}
\def\cp{&\times& \!\!\!}
\def\se{\!\!\! &\simeq& \!\!\!}
\def\hml{\hat{m}_{\ell}}
\def\rr{\hat{r}_{\Lambda}}
\def\ss{\hat{s}}

\renewcommand{\textfraction}{0.2}    
\renewcommand{\topfraction}{0.8}
\renewcommand{\bottomfraction}{0.4}
\renewcommand{\floatpagefraction}{0.8}
\newcommand\mysection{\setcounter{equation}{0}\section}

\def\baeq{\begin{appeq}}     \def\eaeq{\end{appeq}}
\def\baeeq{\begin{appeeq}}   \def\eaeeq{\end{appeeq}}
\newenvironment{appeq}{\beq}{\eeq}
\newenvironment{appeeq}{\beeq}{\eeeq}
\def\bAPP#1#2{
 \markright{APPENDIX #1}
 \addcontentsline{toc}{section}{Appendix #1: #2}
 \medskip
 \medskip
 \begin{center}      {\bf\LARGE Appendix #1 :}{\quad\Large\bf #2}
\end{center}
 \renewcommand{\thesection}{#1.\arabic{section}}
\setcounter{equation}{0}
        \renewcommand{\thehran}{#1.\arabic{hran}}
\renewenvironment{appeq}
  {  \renewcommand{\theequation}{#1.\arabic{equation}}
     \beq
  }{\eeq}
\renewenvironment{appeeq}
  {  \renewcommand{\theequation}{#1.\arabic{equation}}
     \beeq
  }{\eeeq}
\nopagebreak \noindent}

\def\eAPP{\renewcommand{\thehran}{\thesection.\arabic{hran}}}
\renewcommand{\theequation}{\arabic{equation}}
\newcounter{hran}
\renewcommand{\thehran}{\thesection.\arabic{hran}}
\def\bmini{\setcounter{hran}{\value{equation}}
\refstepcounter{hran}\setcounter{equation}{0}
\renewcommand{\theequation}{\thehran\alph{equation}}\begin{eqnarray}}
\def\bminiG#1{\setcounter{hran}{\value{equation}}
\refstepcounter{hran}\setcounter{equation}{-1}
\renewcommand{\theequation}{\thehran\alph{equation}}
\refstepcounter{equation}\label{#1}\begin{eqnarray}}


\newskip\humongous \humongous=0pt plus 1000pt minus 1000pt
\def\caja{\mathsurround=0pt}

\title{\bf Multi-Dimensional Cosmology and GUP}
\author{K. Zeynali$^1$\thanks{Email: k.zeinali@arums.ac.ir}
\hspace{2mm},  \hspace{2mm}
 F. Darabi$^2$\thanks{Email: f.darabi@azaruniv.edu (Corresponding author)   }
 \hspace{2mm}, and \hspace{2mm}
 H. Motavalli$^1$\thanks{Email: motavalli@tabrizu.ac.ir } \\
\centerline{$^1$\small {\em Department of Theoretical Physics
and Astrophysics, University of Tabriz, 51666-16471, Tabriz,
Iran.}}\\
{$^2$\small {\em Department of Physics, Azarbaijan University of Shahid Madani, 53714-161, Tabriz, Iran. }}}

\maketitle
\begin{abstract}
We consider a multidimensional cosmological model with FRW type
metric having 4-dimensional space-time and $d$-dimensional
Ricci-flat internal space sectors with a higher dimensional
cosmological constant. We study the classical cosmology in
commutative and GUP cases and obtain the corresponding exact
solutions for negative and positive cosmological constants. It is
shown that for negative cosmological constant, the commutative and
GUP cases result in finite size universes with smaller size and
longer ages, and larger size and shorter age, respectively. For
positive cosmological constant, the commutative and GUP cases
result in infinite size universes having late time accelerating
behavior in good agreement with current observations. The
accelerating phase starts in the GUP case sooner than the
commutative case. In both commutative and GUP cases, and for both
negative and positive cosmological constants, the internal space
is stabilized to the sub-Planck size, at least within the present
age of the universe. Then, we study the quantum cosmology by
deriving the Wheeler-DeWitt equation, and obtain the exact
solutions in the commutative case and the perturbative solutions
in GUP case, to first order in the GUP small parameter, for both
negative and positive cosmological constants. It is shown that
good correspondence exists between the classical and quantum
solutions.

\end{abstract}

~~~PACS numbers: 98.80.Hw; 04.50.+h

\section{Introduction}
It is well known that quantum description of gravity must be
considered when we want to deal with systems in the Planck scales,
such as very early universe or strong gravitational field of a
black hole. In the absence of gravity, quantum description of a
system can be derived from classical consideration by replacing of
Poisson bracket with usual commutation relations $(\{$ ,
$\}\rightarrow \frac{1}{i\hbar}[$ , $])$. When we want to consider
the gravitational effect in quantum description of a system, some
essential modification in the ordinary quantum principal are
needed. General Uncertainty Principal (GUP) is a modification of
Heisenberg Uncertainty Principal in the Planck scale. Such a
generalization has already been considered in the context of string
theory where the string can not probe distance smaller than the
string size \cite{string1}-\cite{string6}. Some general view to
the GUP and its application to cosmology are proposed in \cite{general GUP1}-\cite{general GUP6} and \cite{general GUP7}, \cite{general GUP8}, respectively.
Michele Maggiore in the discussion of a Gedanken experiment for
the measurement of the area of apparent horizon of a black hole in
quantum gravity , using rather general and model independen
consideration, showed that a minimum length of order of the Planck
length in a Generalized Uncertainty Principal emerges naturally
from any quantum theory of gravity \cite{general GUP3}. In
\cite{general GUP1}, the simplest form of the GUP in a one
dimensional system, is written as:

\bea\label{e1}
 \Delta x \Delta p\geq\frac{\hbar}{2}\Bigg(1+\beta\frac{L_{Pl}^2}{\hbar^2}(\Delta p)^2 \Bigg),
 \eea
 where $L_{Pl}=\sqrt{\frac{G\hbar}{c^3}}=10^{-33}$ is the Planck length and $\beta$ is a positive constant to be of order unity.
 The new term in the above equation is important when $x, \Delta x\approx L_{Pl}$.
 It is possible to show that the above GUP relation (equation \ref{e1}) can be derived from the following generalized
 Heisenberg algebra:
 \bea\label{e2}
  [x,p]=i\hbar (1+\beta p^2).
 \eea

Since cosmology provides the ground for testing physics at high
energy, it seems natural to expect the effects of quantum gravity
in this context. Alternatively, in cosmological systems, the scale
factor, matter field and their conjugate momenta play the role of
dynamical variables of the system; so, introducing GUP in the
corresponding phase space is particularly relevant.

In the past few years the search for a consistent quantum theory
of gravity and the quest for a unification of gravity with other
forces have both led to a renewed interest in theories with extra
spatial dimensions \cite{Duff}-\cite{Schwarz}. Extra spatial
dimensions ideas date back to the seminal work of Kaluza
\cite{Kaluza} and Klein \cite{Klein}. These extra spacial
dimensions must be hidden and assumed to be unseen because they
are compact and have small radius, presumably with typical
dimensions of the order of planck length, $O{(10^{-33}cm)}$. At
Planck time, $t_p=\sqrt{\frac{\hbar G}{c^5}}=O{(10^{-44}s)}$, the
characteristic size of both internal and external dimensions are
likely to have been the same, and the internal dimensions may have
had a more direct role in the dynamics of evolution of the
Universe.

As we discussed above, the existence of extra dimensions and the
influence of GUP become evident at high energy. So, it is clear
that in studying cosmology at high energy, we should presumably
consider extra dimensions and GUP together \cite{Ext1}, \cite{Ext2}.

The paper is organized as follows. In section 2, we construct a
multidimensional cosmology. By convenient coordinate
transformation, we calculate it's Hamiltonian which describes an
isotropic oscillator-ghost-oscillator system. In section 3, we
investigate the equations of motion in classical model, at first
in commutative case and then by considering the influence of GUP.
Section 4, devotes to quantum cosmology. We construct the
Wheeler-DeWitt (WD) equation in commutative and GUP cases and
solve it to obtain the wavefunction of the corresponding
universes.

\section{The Cosmological Model}
It is well known that universe in large scale is homogenous and
isotropic and has 3-dimensional space at least in the order of
experimental tests. So it is possible to exist unseen compact
internal space (Extra dimension) with small radius. For our
investigations, we consider a multi-dimensional cosmological model
in which the space-time is established by a FRW type metric of
4-dimensional space time and a d-dimensional Ricci-flat internal
space \cite{khosravi1}:

\bea\label{e3}
ds^2=-dt^2+\frac{R^2(t)}{(1+\frac{k}{4}r^2)}(dr^2+r^2d\Omega^2)+a^2(t)g_{ij}^{(d)}dx^idx^j,
  \eea
where $k=1, 0, -1$ represents the usual spatial curvature of
external space in FRW model, $R(t)$ and $a(t)$ are the scale
factors of the external and internal space respectively, and
$g_{ij}^{(d)}$ is the metric of the internal space which is
assumed to be Ricci-flat. The total number of dimensions is
$D=3+d$. The Ricci scalar can be derived from the metric(\ref{e3})
\cite{khosravi1}
  \bea\label{e4}
  {\cal R}=6\Bigg(\frac{\ddot{ R}}{ R}+\frac{k+\ddot{ R}^2}{R^2}\Bigg)
  +2d\frac{\ddot{a}}{a}+d(d-1)\Bigg(\frac{\dot{a}}{a}\Bigg)^2+6d\frac{\dot{a}\dot{R}}{a R},
  \eea
where a dot represents differentiation with respect to $t$. We
consider an Einstein-Hilbert action functional with a
$D$-dimensional cosmological constant $\Lambda$:
  \bea\label{e6}
  {\cal S}=\frac{1}{2k^2_D}\int_M
  d^Dx\sqrt{-g}({\cal R}-2\Lambda)+{\cal S}_{YGH},
  \eea
where $k_D$ is the $D$-dimensional gravitational constant and
${\cal S}_{YGH}$ is the York-Gibbons-Hawking boundary term.
Substitution Eq.(\ref{e4}), after dimensional reduction we have:
  \bea\label{e7}
  {\cal S}=-v_{D-1}\int dt\Bigg \{6\dot{ R}^2\Phi R+6\dot{ R}\dot{\Phi}
  R^2+\frac{d-1}{d}\frac{\dot{\Phi}^2}{\Phi} R^3-6k\Phi R+2\Phi R^3\Lambda \Bigg\},
  \eea

  where
  \bea\label{e8}   \Phi=\Bigg(\frac{a}{a_0}\Bigg)^d,
  \eea
and $a_0$ is the compactification scale of the internal space at
present time. We can set $v_{D-1}=1$. To make the Lagrangian
suitable, we consider the following change of
variables\footnote{Here we have some differences from
\cite{khosravi1} in defining the parameters of equations
(\ref{e11}).}.
  \bea\label{e9}
   \Phi  R^3=\Upsilon^2(x_1^2-x_2^2),
  \eea
  \bea\label{e10}
   \Phi^{\rho_+}  R^{\sigma_-}&=&\Upsilon(x_1+x_2),\nnb \\
   \Phi^{\rho_-}  R^{\sigma_+}&=&\Upsilon(x_1-x_2).
  \eea
with
   \bea\label{e11}
   \rho_\pm
   &=&\frac{1}{2}\pm\frac{1}{2}\sqrt{\frac{3}{d(d+2)}},
\nnb \\
   \sigma_\pm
   &=&\frac{3}{2}\pm\frac{1}{2}\sqrt{\frac{3d}{d+2}},\nnb \\
   \Upsilon&=&\frac{1}{2}\sqrt{\frac{d+3}{d+2}}.
  \eea
where $R=R(x_1, x_2)$ and $\Phi=\Phi(x_1, x_2)$ are functions of
new variables $x_1, x_2$. Using the above transformations and
concentrating on $k=0$, the Lagrangian becomes
  \bea\label{e12}
   {\cal L}=
   (\dot{x_1}^2-\dot{x_2}^2)+\frac{\Lambda}{2}\Bigg(\frac{d+3}{d+2}\Bigg)\Big(x_1^2-x_2^2\Big).
  \eea
We can write the effective Hamiltonian as \footnote{ In the Hamiltonian
of cosmological models having matter contents, usually a wrong sign between geometric and matter sectors appears. This is due to the fact that the matter content of the universe has positive energy while the gravitational content has negative energy. In the present model, there is no explicit matter
content, however, we may interpret the sector of extra compact dimensions
as a fictious matter field. Then, the Hamiltonian of the variables $R$, $a$
having a sign difference between the four and extra dimensional sectors
is transformed to that of the variables  $x_1, x_2$ with the same sign difference
between $x_1$ and $x_2$ parts. These Hamiltonians are usualy called oscillator-ghost-oscillator.}
  \bea\label{e13}
   {\cal H}=\left(\frac{p_1^2}{4}+\omega^2x_1^2\right)-\left(\frac{p_2^2}{4}+\omega^2
   x_2^2\right),
  \eea
where
  \bea\label{e14}
   \omega^2=-\frac{1}{2}\Bigg(\frac{d+3}{d+2}\Bigg)\Lambda.
  \eea

Equations (\ref{e13}), (\ref{e14}) and (\ref{e9}) show that the
potential energy for our oscillator-ghost-oscillator system
\cite{Matej Pavsic} is proportional to vacuum energy $\Lambda$
times the volume of the multidimensional universe and the total
momentum of such a system is
$p^2_{tot}=\frac{p_1^2}{2}-\frac{p_2^2}{2}$.

The effective oscillators in Eq.(\ref{e13}) appear to be
decoupled and the dynamics of two dynamical variables, namely $x_1$ and $x_2$ seems to be independent of each other. However, the Hamiltonian constraint $({\cal H}=0)$ imposing on oscillators connects their dynamics \footnote{We know that general relativity is a {\it time reparametrization invariant} theory. Every theory which is diffeomorphism invariant casts into the
constraint systems. Therefore, general relativity is a constraint
system whose constraint is the zero energy condition $({\cal
H}=0)$.}. Therefore, the dynamics of external large dimensions $R$ and internal compactified extra dimensions $a$ are inevitably related to each other by the equation (\ref{e10}).

\section{Classical solutions}
\subsection{Commutative case}
 As in \cite{khosravi1, Matej Pavsic}, the dynamical variable defined in (\ref{e10}) and their conjugate momenta
 satisfy:
 \bea\label{e15}
   \{x_\mu, p_\nu \}_P=\eta_{\mu\nu},
  \eea
where $\eta_{\mu\nu}$ is the two dimensional Minkowski metric and
$\{$ , $\}_P$ represents the Poisson bracket. The equations of
motion can be written as:
\bea\label{e17}  \dot{x}_\mu&=&\{x_\mu,{\cal H}\}_P=\frac{1}{2}p_\mu,\nnb \\
\dot{p}_\mu&=&\{p_\mu,{\cal H}\}_P=-2\omega^2x_\mu ,
 \eea
Combining two equations in (\ref{e17}) we obtain
 \bea\label{e18}
\ddot{x_\mu}+\omega^2 x_\mu=0.
 \eea

First, we assume a negative cosmological constant. According to
(\ref{e14}), $\omega^2$ is then positive. Considering (\ref{e13}),
it is obvious that Eq.(\ref{e18}) describes the equations of two
ordinary uncoupled harmonic oscillators whose solutions are

\bea\label{e19} x_\mu(t)=A_\mu e^{i\omega t}+B_\mu e^{-i\omega t},
 \eea

where $A_\mu$ and $B_\mu$ are constants of integration. Imposing
the Hamiltonian constraint $({\cal H}=0)$ introduce the following
relation on these constants.
 \bea\label{e20} A_\mu B^\mu=0.
 \eea
Finally, using (\ref{e8}) and (\ref{e10}), the scale factors take the following forms
 \bea\label{e21}
a(t)&=&k_1[\sin(\omega t+\phi_1)]^{\frac{\sigma_+}{d(\rho_+ \sigma_+ -\rho_- \sigma_-)}}[\sin(\omega t+\phi_2)]^{\frac{-\sigma_-}{d(\rho_+ \sigma_+ -\rho_- \sigma_-)}},
 \nnb \\
R(t)&=&k_2[\sin(\omega t+\phi_1)]^{\frac{-\rho_-}{\rho_+ \sigma_+ -\rho_- \sigma_-}}[\sin(\omega t+\phi_2)]^{\frac{\rho_+}{\rho_+ \sigma_+ -\rho_- \sigma_-}},
 \eea
where $k_1$ and $k_2$ are arbitrary constants and $\phi_1$ and $\phi_2$ are arbitrary phases.
If we consider the Hamiltonian constraint, the following relation is imposed on these constants
 \bea\label{e22}
\frac{4(d+2)}{d+3}k_1^d k_2^3\cos(\phi_1-\phi_2)=0.
 \eea
Because of $k_1$, $k_2\neq0$, this leads to
$\phi_1-\phi_2=\frac{\pi}{2}$. In what follows, we investigate the
behavior of a universe with one internal dimension $(D=3+1)$.

By setting $\phi_1=\frac{\pi}{2}$ and $\phi_2=0$, we find:

 \bea\label{e23}
R(t)&=&k_2\sqrt{\sin(\omega t)},\nnb \\
a(t)&=&k_1\frac{\cos(\omega t)}{\sqrt{\sin(\omega t)}},
 \eea
 whose classical loci is given by
 \bea\label{e2-3}
a=\frac{1}{R}\sqrt{1-R^4},
 \eea
where the values of $k_1$ and $k_2$ are absorbed in the values of $R$ and $a$. Figure.2 (left), shows this loci for negative cosmological constant.
It is seen that the dynamics of $R$ is constrained by the dynamics of $a$
and vice versa. This is expectable, because the dynamics of two variables
is subject to the zero energy condition so called Hamiltonian constraint. 
Using these solutions, we can calculate the Hubble and
deceleration parameter for both $R(t)$ and $a(t)$ as
\bea\label{e24}
H_R(t)&=&\frac{\dot{R}(t)}{R(t)}=\frac{\omega}{2}\cot(\omega t),
\nnb \\
q_R(t)&=&- \frac{R(t)\ddot{R}(t)}{\dot{R}^2(t)}=1+2\tan^2(\omega
t),
\nnb \\
H_a(t)&=&\frac{\dot{a}(t)}{a(t)}=-\frac{\omega}{2}(\cot(\omega
t)+2\tan(\omega t)),
\nnb \\
q_a(t)&=&-
\frac{a(t)\ddot{a}(t)}{\dot{a}^2(t)}=-\frac{2\cos^2(\omega
t)(5+\cos(2\omega t))}{(-3+\cos(2\omega t))^2}.
 \eea

From (\ref{e23}), we see that as time increases from $t=0$ toward
$\frac{\pi}{2\omega}$, $R(t)$ and $a(t)$ are increasing and
decreasing functions of $t$, toward the maximum and zero, respectively. At $t=\frac{\pi}{2\omega}$, $R(t)$ reaches to its maximum and $a(t)$ is contracted toward zero. From $t=\frac{\pi}{2\omega}$ to $t=\frac{\pi}{\omega}$, $R(t)$ is contracted toward zero while $a(t)$ is decreasing in the region of negative values. Although the negative value for $a(t)$
is meaningless, but it is $a^2(t)>0$ which is physically viable in the definition
of metric\footnote{This situation is similar to that of wavefunction in quantum
mechanics, where it is the square of the wavefunction which is physical and not the wavefunction itself.}. Therefore, we stick to the behavior of
$a^2(t)>0$ instead of $a(t)$. As is seen in Fig.1 (solid lines), from $t=\frac{\pi}{2\omega}$ to $t=\frac{\pi}{\omega}$, both $R^2(t)$ and $a^2(t)$ become decreasing and increasing functions, respectively. According to the behavior mentioned above, the four dimensional sector begins from a Big Bang at $t\simeq0$, then expands till $t=\frac{\pi}{2\omega}$ toward a maximum value after which the gravity takes over the expansion and this sector starts contracting toward a big crunch at $t\simeq\frac{\pi}{\omega}$. In order to avoid of initial singularity, we will assume the Big Bang to occur at the Planck time $t_{Pl}$. In fact, it is reasonable that at planck time we set $R(t_{Pl})=a(t_{Pl})$. In this regard, the extra dimensional sector begins from $a^2(t_{Pl})$, then contracts till $t=\frac{\pi}{2\omega}$ toward a vanishing minimum value after which this sector starts expanding toward $a^2(\frac{\pi}{\omega}-t_{Pl})=a^2(t_{Pl})$.
we see that during the whole time evolution of universe ($t_{Pl}\leq t\leq
\frac{\pi}{\omega}-t_{Pl})$, the scale factor of internal space is
contracted toward zero and can never exceed $a(t_{Pl})$.
By considering present value of Hubble constant, we see that the
age of universe is $t_{present}=\frac{1}{\omega}
\cot^{-1}(\frac{2H_0}{\omega})$
whose order of magnitude is in agreement with present observations, namely
$\omega^{-1}\approx10^{17}s$. So, the present universe is in tideway to get to the maximum and minimum of $R^2(t)$ and $a^2(t)$, respectively, within $\Delta t\approx 0.57\omega^{-1}$. 

In the time interval $\frac{\pi}{\omega}\leq
t\leq \frac{2\pi}{\omega}$, $R^2(t)$ and $a^2(t)$ become negative
Figs.1 (solid lines). To avoid this problem we may consider two
options:

 1) Classically, negative values for $R^2(t)$ and $a^2(t)$ are nonphysical,
 so we may leave the cosmology with imaginary scale factors as the nonphysical
 solutions of the Einstein equations and think that the physical universe will end at $t=\frac{\pi}{\omega}$, with no further extension or history.

 2) Quantum mechanically, as the scale factor $R(t)$ goes to zero (when $t$ approaches the big crunch singularity),
 the strong quantum gravitational effects can drastically change the initial conditions so that$R^2(t)$ and $a^2(t)$
 may become again nonzero positive functions capable of establishing another Big Bang.

By, considering \bea
R(t_{Pl})=k_2\sqrt{\sin(\omega t_{Pl})},\nnb \\
a(t_{Pl})=k_1\frac{\cos(\omega t_{Pl})}{\sqrt{\sin(\omega
t_{Pl})}},
\eea
the initial condition $R(t_{Pl})=a(t_{Pl})$ results in
\bea
\frac{k_2}{k_1}=10^{61},
\eea
by which we obtain the following ratio
\bea
\frac{R(t)}{a(t)}=10^{61}\tan(\omega t).
\eea
If we consider the present radius of external space to be equal to the radius
of observed universe, namely $10^{28}cm$, then we see that the present radius of internal space is about the Planck length $(10^{-33}cm)$.

By considering smallness of the internal dimension and using (\ref{e24}) to evaluate the approximate magnitude of $H_a(t_{present})\sim \omega \approx10^{-17}$,
we see that the variation of $a(t)$ at present time is ignorable. So, in the context of negative cosmological constant, the compactification and the stabilization of
the internal space at present status of the universe is established.

\begin{figure}
\vskip 1.5 cm
    \includegraphics{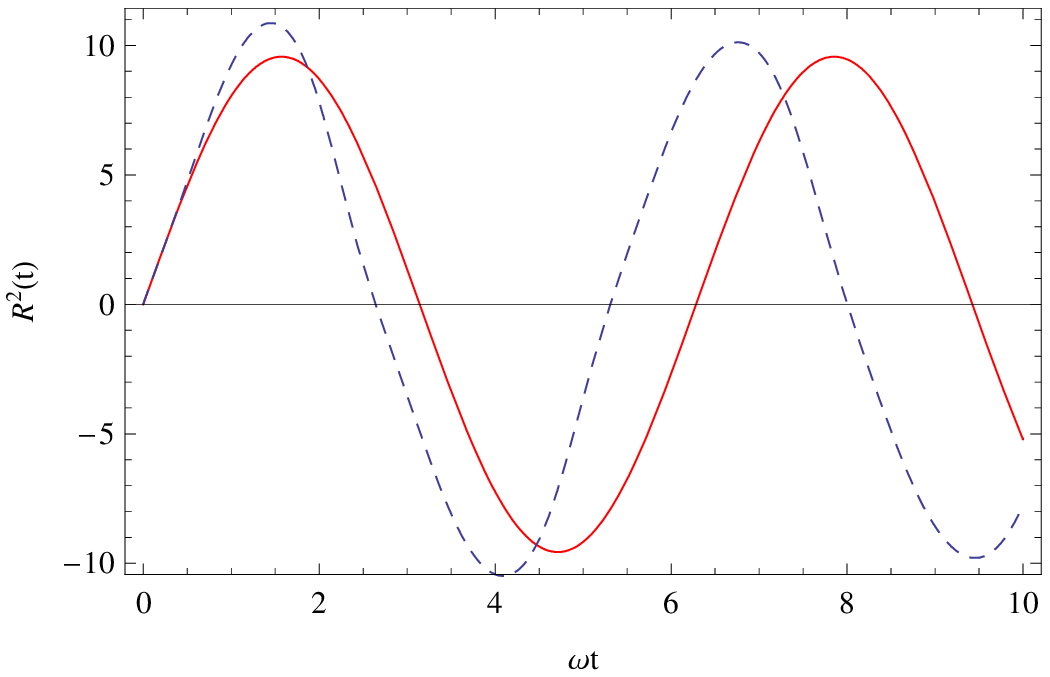}
    \includegraphics{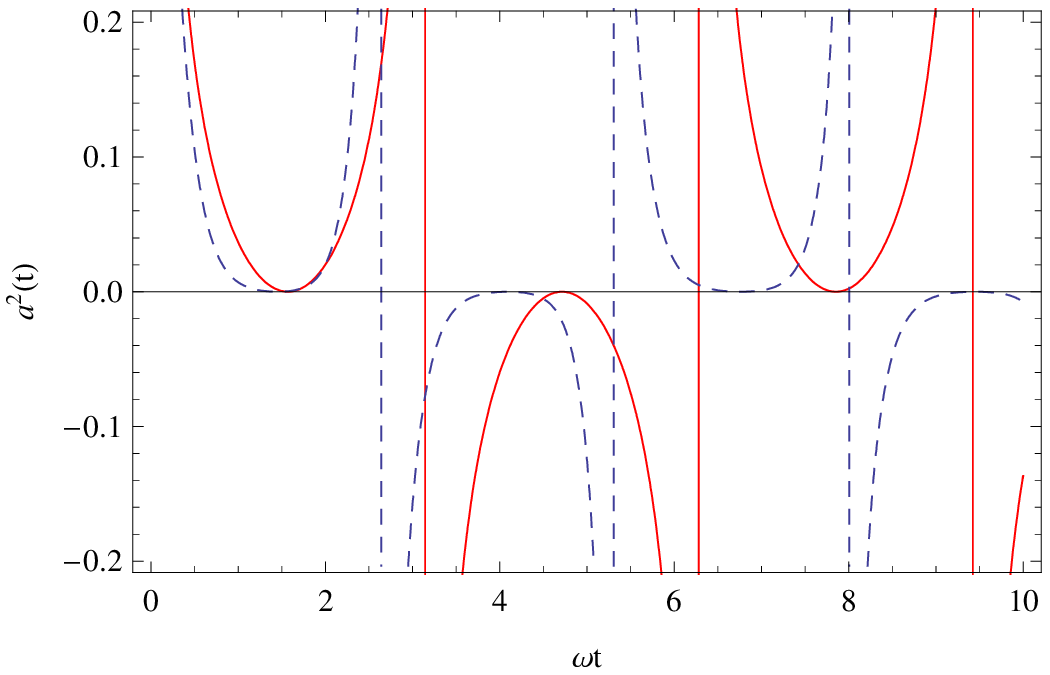}
\vskip 2.5cm \caption{Time evolution of the (squared) scale factors of
universe with one extra dimension and negative cosmological
constant. Solid and dashed lines refer to the scale factors in
commutative and GUP framework respectively. Left and
right figures are the external and internal dimensions
respectively. }
\end{figure}

\begin{figure}
\vskip 2.5 cm
    \includegraphics{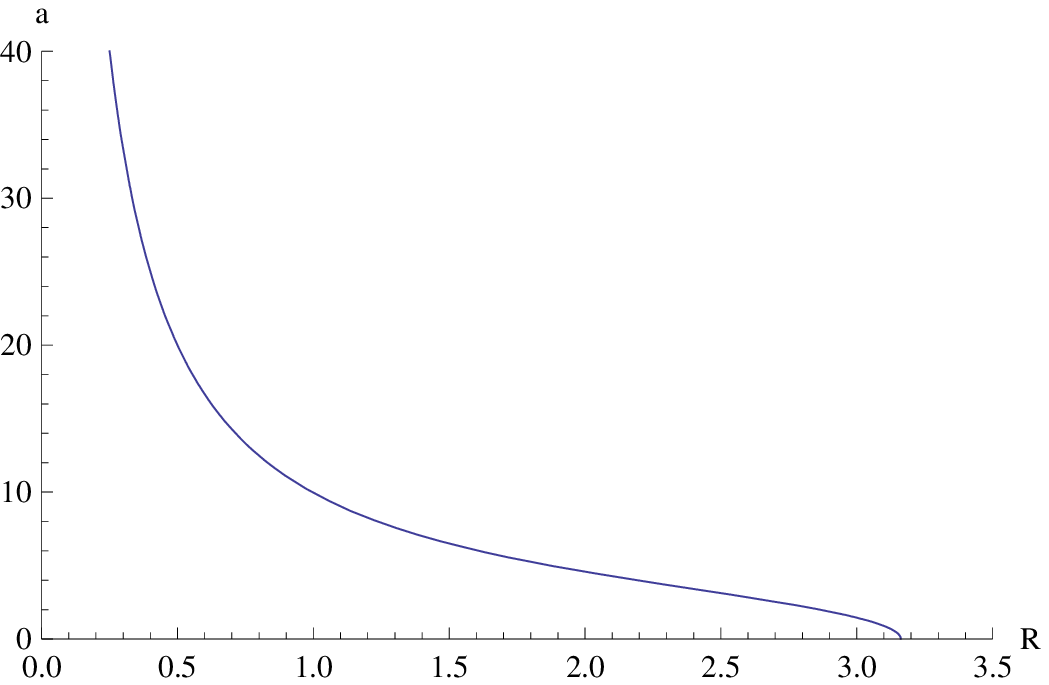}
    \includegraphics{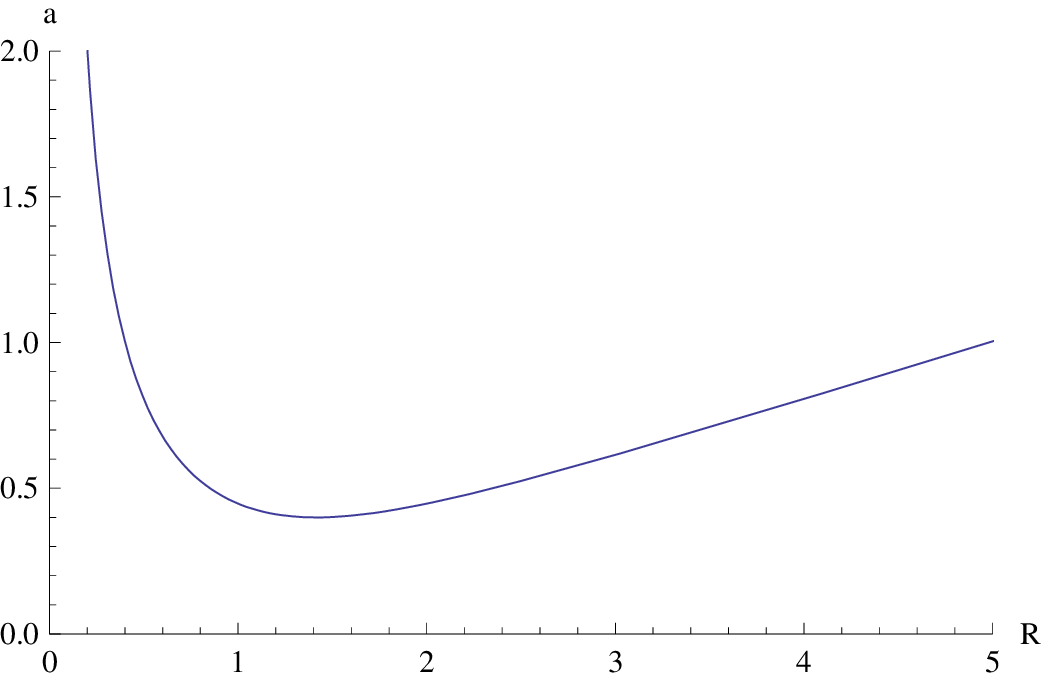}
\vskip 2.5cm \caption{Left figure is classical loci in the
negative cosmological context and right figure is the classical
loci in the positive cosmological context. }
\end{figure}

Now, we assume a positive cosmological constant for which
$\omega^2$ is negative. If we replace $\omega^2$ with
$-\omega^2$ in Eq.(\ref{e18}), the new solutions are then obtained by
replacing trigonometric functions by their hyperbolic counterparts
in the solutions (\ref{e21}). By considering Hamiltonian constrain we obtain
the following solutions
 \bea\label{e26}
a(t)&=&k_1[\cosh(\omega t)]^{\frac{\sigma_+}{d(\rho_+ \sigma_+ -\rho_- \sigma_-)}}[\sinh(\omega t)]^{\frac{-\sigma_-}{d(\rho_+ \sigma_+ -\rho_- \sigma_-)}}, \nnb \\
R(t)&=&k_2[\cosh(\omega t)]^{\frac{-\rho_-}{\rho_+ \sigma_+ -\rho_- \sigma_-}}[\sinh(\omega t)]^{\frac{\rho_+}{\rho_+ \sigma_+ -\rho_- \sigma_-}},
 \eea
 where for $d=1$ we have
\bea\label{e27}
a(t)&=&k_1\frac{\cosh(\omega t)}{\sqrt{\sinh(\omega t)}}, \nnb \\
R(t)&=&k_2\sqrt{\sinh(\omega t)},
 \eea
 whose classical loci is given by
 \bea\label{e2-3'}
a=\frac{1}{R}\sqrt{1+R^4}.
 \eea
Figure.2 (right), shows this loci for positive cosmological constant.
As in the negative cosmological constant case, the dynamics of both $R$ and $a$ is constrained by the Hamiltonian constraint.
The scale factor ratio is obtained
\bea \frac{R(t)}{a(t)}=10^{61}\tanh(\omega t), \eea 
and we have
 \bea\label{e28}
H_R(t)&=&\frac{\dot{R}(t)}{R(t)}=\frac{\omega}{2}\coth(\omega t),
\nnb \\
q_R(t)&=&- \frac{R(t)\ddot{R}(t)}{\dot{R}^2(t)}=1-2\tanh^2(\omega t),
\nnb \\
H_a(t)&=&\frac{\dot{a}(t)}{a(t)}=\frac{\omega}{2}(-\coth(\omega t)+2\tanh(\omega t)),
\nnb \\
q_a(t)&=&- \frac{a(t)\ddot{a}(t)}{\dot{a}^2(t)}=-\frac{2\cosh^2(\omega t)(5+\cosh(2\omega t))}{(-3+\cosh(2\omega t))^2}.
 \eea
Note that, like the case of negative cosmological constant, the magnitude of the radius of external to internal space is asymptotically ($t\rightarrow
\infty$) about $ 10^{61}$.
Equations (\ref{e27}) show that $R(t)$ is an increasing function of time. But, $a(t)$ at first decrease with time till $t\simeq0.88\omega^{-1}$ and then increase exponentially (see Fig. 3). As we mentioned above, at planck time, the characteristic size of both internal and external dimensions are
assumed to be the same, namely $(R(t_{Pl})=a(t_{Pl}))$. If we
consider the age of universe about $\omega^{-1}\simeq10^{17}s$, we
see that at present time we are around the minimum point of $a(t)$.
Also by some calculation on $a(t)$, we find that in the time interval
$t_{Pl}\leq t \leq 141\omega^{-1}$, $a(t)$ can never exceed it's initial value at Planck time $a(t_{Pl})$. Therefore, the internal scale factor remains small for a very long period with a duration of about 140 times of the present age of the universe.

\begin{figure}
\vskip 1.5 cm
    \includegraphics{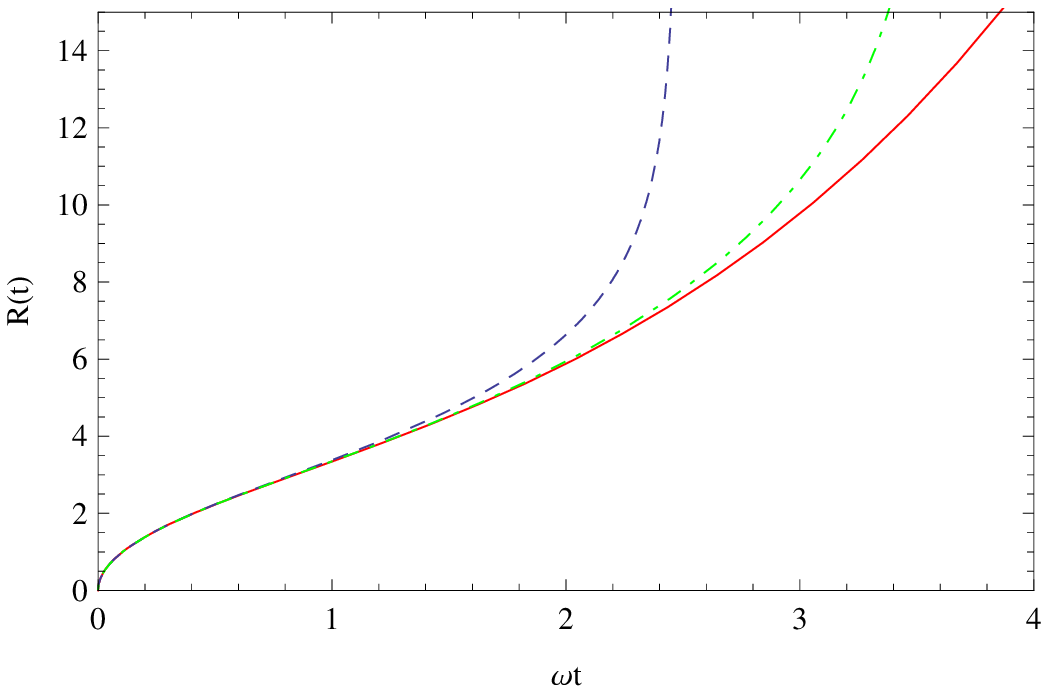}
    \includegraphics{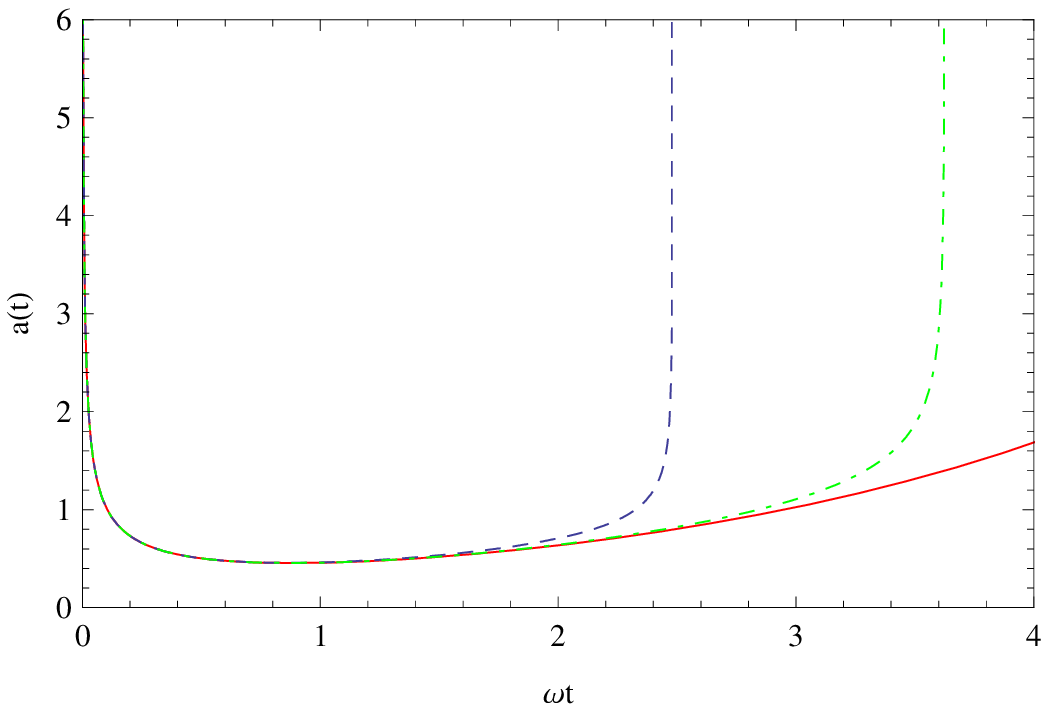}
\vskip 2.5cm \caption{Time evolution of the scale factors of
universe with one extra dimension and positive cosmological
constant. Solid lines refer to the scale factors in commutative case, dashed and dot-dashed lines refer to the scale factors in GUP framework with $\beta=10^{-4}$ and $\beta=10^{-5}$ respectively. Left and right figures are the
external and internal dimensions respectively.}
\end{figure}

\newpage

\begin{figure}
\vskip 2.5 cm
    \includegraphics{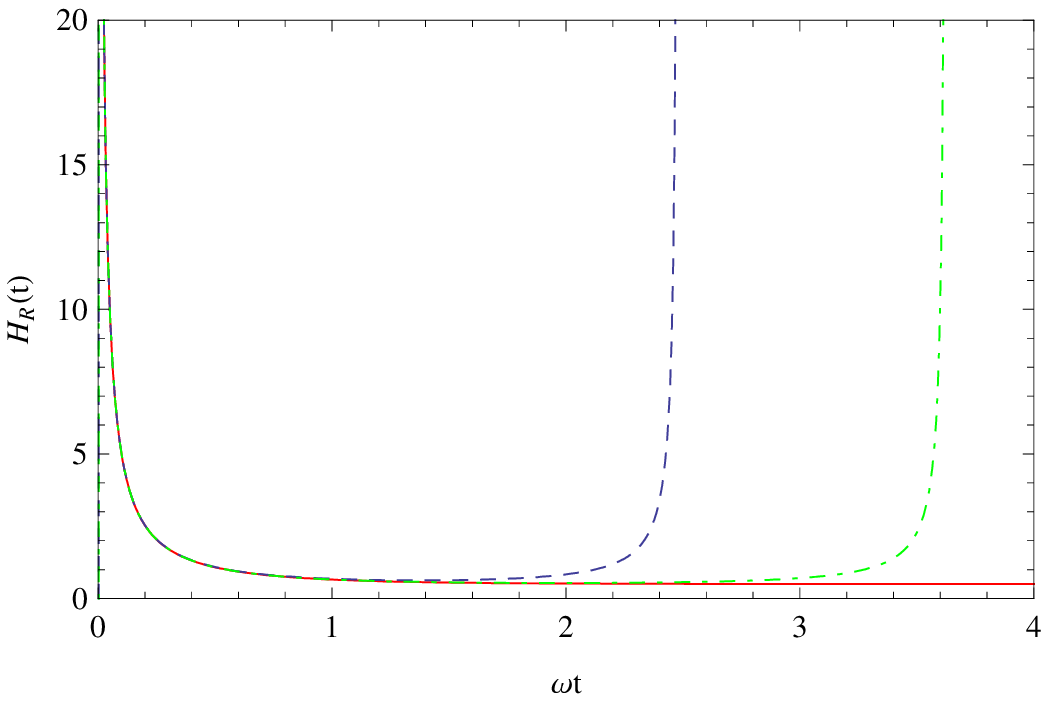}
    \includegraphics{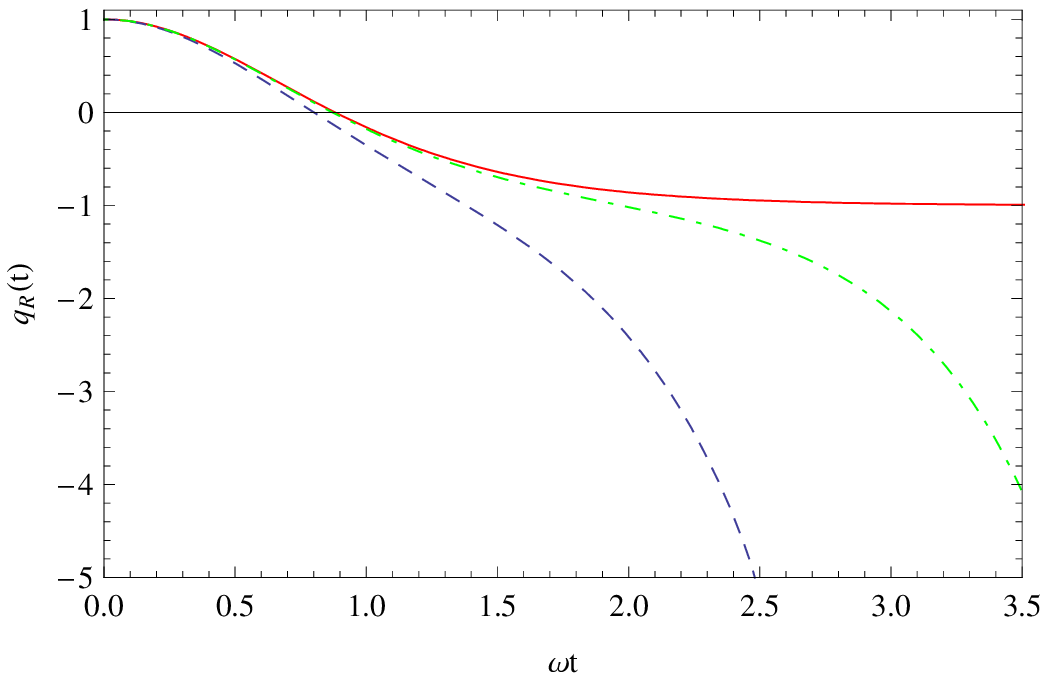}
\vskip 2.5cm \caption{Left and right figures are respectively
Hubble an deceleration parameters of external space for universe
with one extra dimension and positive cosmological constant.
Solid lines refer to the scale factors in commutative case, dashed and dot-dashed lines refer to the scale factors in GUP framework with $\beta=10^{-4}$ and $\beta=10^{-5}$ respectively.}
\end{figure}

\subsection{GUP case}

In more than one dimension, it can be shown that the generalized
Heisenberg algebra corresponding to GUP is defined by the following
commutation relations \cite{general GUP1, sepangi}

\bea\label{e29}
 [x_i,p_j]=i(\delta_{ij}+\beta\delta_{ij}p^2+\beta'p_ip_j),
 \eea
where $p^2=\sum p_i p_i$ and $\beta \beta'>0$ are considered as
small quantities of first order. Throughout the whole process we
work in the units with $\hbar=G=c=1$. We assume that momenta commute
with momenta 
 \bea\label{e30} [p_i,p_j]=0.
 \eea
Using Jacobi identity
$[[x_i,x_j],p_k]+[[x_j,p_k],x_i]+[[p_k,x_i],x_j]=0$ the
commutation relation for the coordinates are obtained as:
\bea\label{e31}
[x_i,x_j]=i\frac{(2\beta-\beta')+(2\beta+\beta')\beta p^2}
{(1+\beta p^2)}(p_ix_j-p_jx_i).
 \eea

\begin{figure}
\vskip 1.5 cm
    \includegraphics{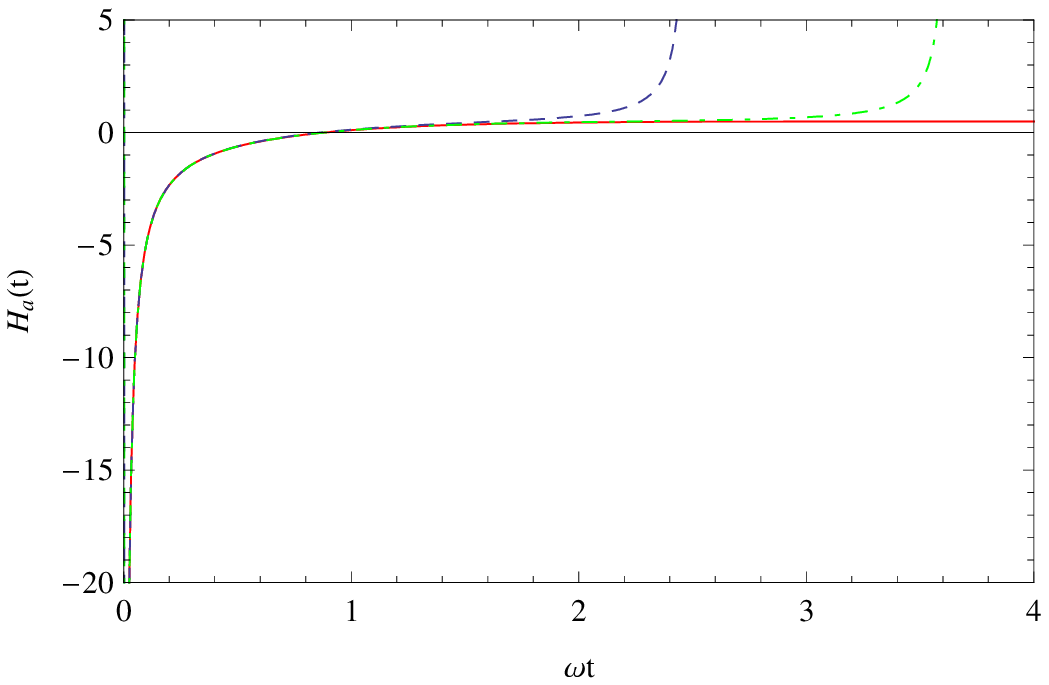}
    \includegraphics{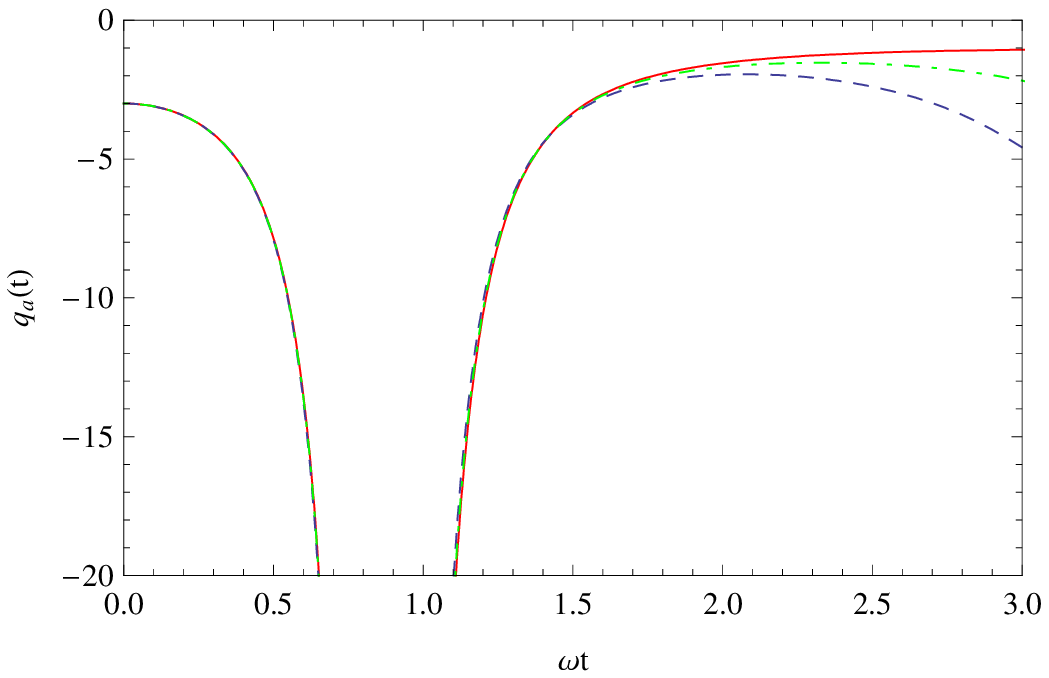}
\vskip 2.5cm \caption{Left and right figures are respectively
hubble an deceleration parameters of internal space for universe
with one extra dimension and positive cosmological constant.
Solid lines refer to the scale factors in commutative case, dashed and dot-dashed lines refer to the scale factors in GUP framework with $\beta=10^{-4}$ and $\beta=10^{-5}$ respectively.}
\end{figure}
It is clear from the above equation that the coordinates do not
commute. So, we can not work in the position space to construct
Hilbert space representations. But, if we choose the special case
$\beta'=2\beta$, we can see from equation (\ref{e31}) that
coordinates commute to first order in $\beta$ and thus we can work
in coordinate representation. By the following definitions,
equations (\ref{e29}) and (\ref{e30}) with $\beta'=2\beta$ can be
realized up to first order in $\beta$ (see
Appendix)\cite{Majumder}

\bea\label{e31.2} x_i=x_{i0}, &    &p_i=p_{i0}(1+\beta p_0^2),
 \eea
where $[x_{i0}, p_{j0}]=i \delta_{ij}$, $p_0^2=\sum p_{i0} p_{i0}$
and $p_{i0}=-i\frac{\partial}{\partial x_{i0}}$. We can show that
the $p^2$ term in any Hamiltonian can be written as
\cite{Majumder}

 \bea\label{e32}
p^2=p_0^2+2\beta p_0^4.
 \eea

In our model, by considering $\beta'=2\beta$ and in first order in
$\beta$, the commutation relation between position and momentum
operators can be summarized as

\bea\label{e33} [x_1,p_1]=i(1+\beta p^2+2\beta p_1^2),&
& [x_2,p_2]=i(1+\beta p^2+2\beta p_2^2),
 \eea

\bea\label{e34}   [x_1,p_2]=[x_2,p_1]=2i\beta p_1p_2,
 \eea

 \bea\label{e35}  [x_i,x_j]=[p_i,p_j]=0, &     &   i,j=1,2.
 \eea
where $p^2_{tot}=\frac{1}{2}(p_1^2-p_2^2)$. We aim to investigate
the effects of the classical version of GUP. As we know, we must
replace the quantum mechanical commutators with the classical
poisson bracket as $[P,Q]\rightarrow i\{P,Q\}$. Thus, in classical
phase space the GUP deformed poisson algebra is achieved from
(\ref{e33})-(\ref{e35})
 by considering $[P,Q]\rightarrow i\{P,Q\}$.

\bea\label{e36} \{x_1,p_1\}=(1+\beta p^2+2\beta p_1^2),& &
\{x_2,p_2\}=(1+\beta p^2+2\beta p_2^2),
 \eea

\bea\label{e37}   \{x_1,p_2\}=\{x_2,p_1\}=2\beta p_1p_2,
 \eea

 \bea\label{e38}  \{x_i,x_j\}=\{p_i,p_j\}=0,    i,j=1,2,
 \eea
In \cite{S. Benczik} such a deformations algebra is used on
classical orbit of particles in a central force field and on
kepler third law. It is notable that this modification is
significant at Plank scale and we need quantum description. But,
before quantizing the model we want to construct a deformed
classical cosmology. Note that, in transition from quantum
commutation relations to Poisson brackets we Shall keep the GUP's
parameter $\beta$ fixed as $\hbar\rightarrow 0$. The equations of
motion can be written as (see \cite{Lay})

\bea\label{e39}  \dot{x}_1&=&\{x_1,H\}=\frac{1}{2}p_1(1+5\beta
p^2), \eea

\bea\label{e40}  \dot{x}_2&=&\{x_2,H\}=-\frac{1}{2}p_2(1-3\beta
p^2), \eea
 \bea\label{e41}  \dot{p}_1&=&\{p_1,H\}=-2\omega^2x_1(1+\beta p^2+2\beta p_1^2)
 +4\omega^2\beta x_2p_1p_2,
\eea
\bea\label{e42}  \dot{p}_2&=&\{p_2,H\}=2\omega^2x_2(1+\beta
p^2+2\beta p_2^2)
 -4\omega^2\beta x_1p_1p_2.
\eea

These show that the deformed classical cosmology forms a system of
nonlinear coupled differential equations that can not be solved
analytically.

\begin{figure}
\vskip 1.5 cm
    \includegraphics{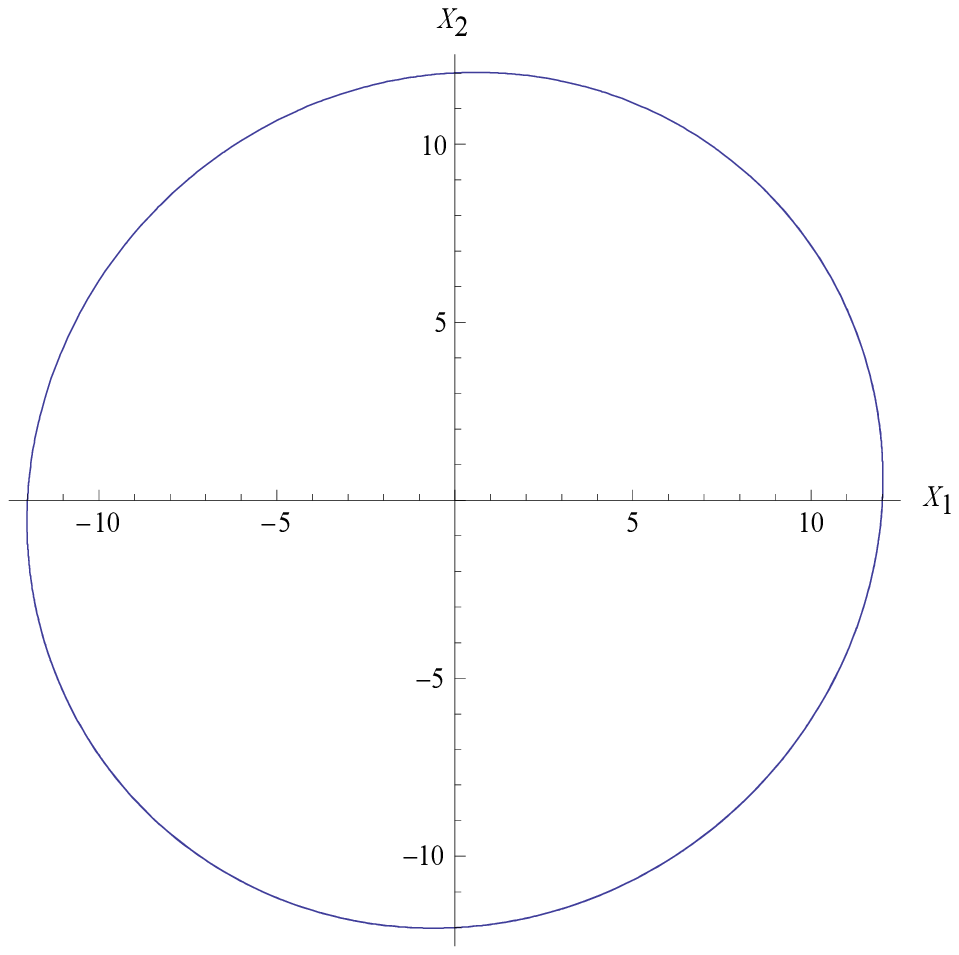}
    \includegraphics{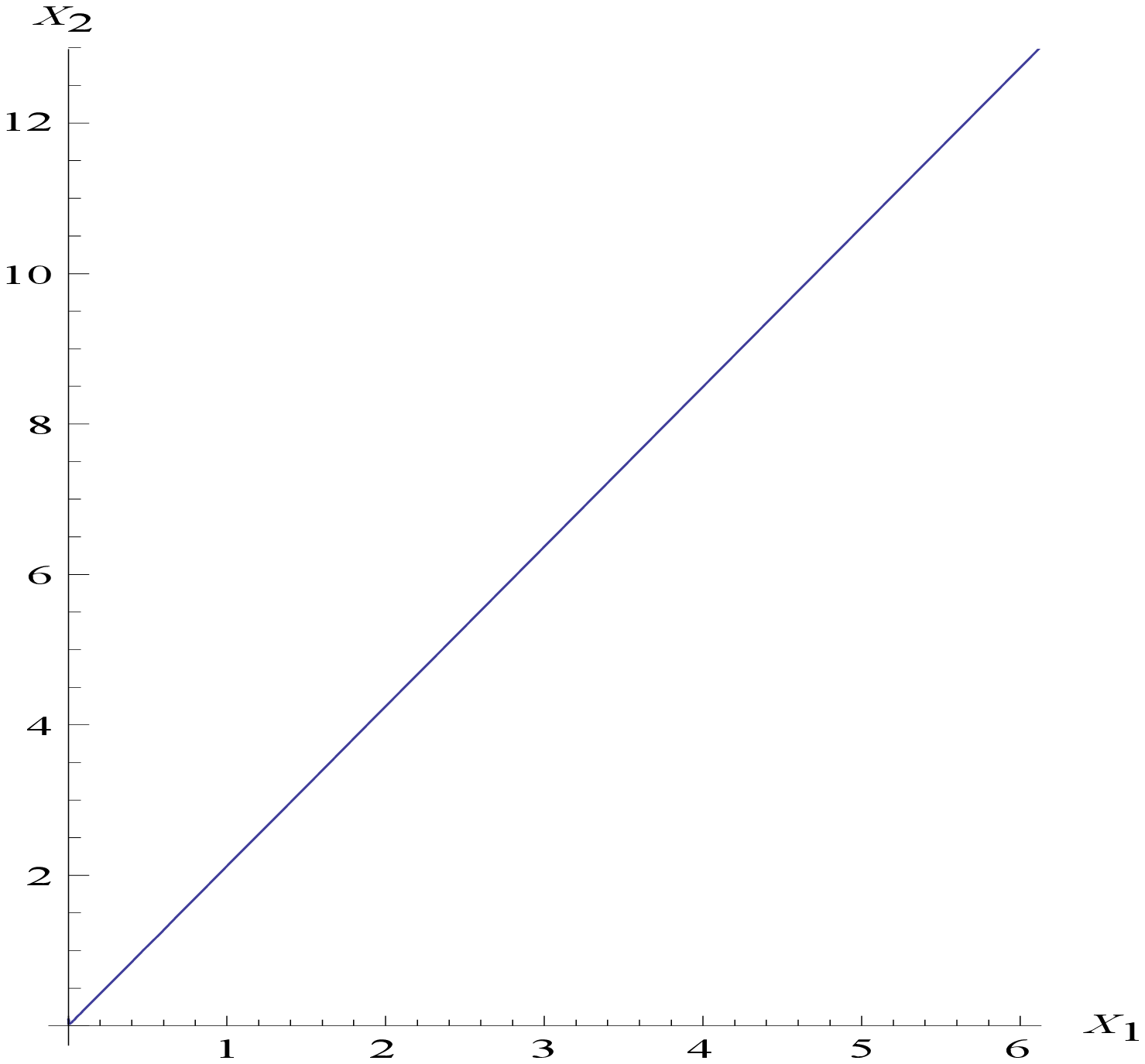}
\vskip 2.5cm \caption{Left and right figures show the classical
trajectories in $ x_1-x_2 $ plane for a universe with one extra
dimension in the context of negative and positive cosmological
constant, respectively.}
\end{figure}

 In the context of negative cosmological constant, $\omega^2$ is positive
(\ref{e14}). Numerical solutions of Eqs. (\ref{e39})-(\ref{e42}) shows that $R(t)$ and $a(t)$ like
 the classical ones have periodic behavior, but the time interval between Big Bang and big crunch
 is shortened (dashed lines, Fig.1).
For later period, the time interval becomes longer with respect to
first one while the maximum value of $R(t)$ is shorter. Comparison
of the results in the present GUP framework with the previous
classical commutative case (Fig. 1) reveals that in the present
framework we have larger external and smaller internal dimensions.
Also the rate of variation of $a(t)$ is less than the previous
case. So, the compactification and the stabilization of internal
space in GUP framework is more favored than the classical
commutative case.

By replacing $\omega^2$ with $-\omega^2$ in Eqs.(\ref{e39})-(\ref{e42}),
we can get the corresponding equations in the case of positive cosmological constant. By numerical analysis, we find that at early time the scale factors behave like classical commutative case (Fig. 3), but at late time both the external and internal dimensions are greater than those of commutative case.
Also, investigation in time dependence of the Hubble and deceleration parameters
shows that at early time these parameters nearly behave like classical commutative
case, but at late time they behave very differently. 

Some recent observations indicate that the universe is currently
undergoing an accelerating period of expansion. This is consistent
with the results obtained here considering a positive cosmological
constant. Depiction of the Hubble and deceleration parameteres for
the scale factors $R$ and $a$ can be of useful help to understand
the origin of current acceleration in this multidimensional model.
To this end, we have depicted $H_R$, $q_R$ and $H_a$, $q_a$ within
the figures 4 and 5 (see solid lines). Figure 4 shows that $q_R$
becomes negative a little bit earlier than $\omega t \sim1$,
namely the present age of the universe. This means, the
acceleration of the universe has started recently in the present
multidimensional commutative case. On the other hand, Fig.5 shows
that $q_a$ is always negative and has a minimum at the position
where $q_R$ becomes negative. Therefore, it seems the behavior of
$q_a$ is responsible for the behavior of $q_R$ and vice versa, as
is interpreted in the following. The present model describes a
multidimensional cosmology, with a positive cosmological constant.
Typically, a standard 4-dimensional FRW cosmology with a positive
cosmological constant predicts an accelerating behavior of the
scale factor. In a multidimensional cosmology, however, it is
reasonable to think that the overall repulsive force due to the
positive cosmological constant manifests as an interplay between
the accelerating and decelerating behaviors of the external and
internal scale factors. Looking at the figures 4 and 5 reveals
that, typically for both commutative and GUP cases, at the
beginning of time, $q_R$ is positive ($R$ is decelerating) while
$q_a$ is negative ($a$ is accelerating). As time passes, $q_R$
approaches the threshold of negative values ($R$ is less
decelerating ) while $q_a$ goes to more negative values ($a$ is
highly accelerating). When $q_R$ enters the region of negative
values ($R$ is accelerating) the $q_a$ reaches its minimum ($a$
stops the increasing acceleration; the minimum is not shown in
Fig.5). Finally, $q_R$ becomes more negative ($R$ is highly
accelerating) and $q_a$ goes to rather less negative values ($a$
is slowly accelerating). In order to better compare the results obtained in commutative and GUP cases, we have depicted all the GUP diagrams for two values of the parameter $\beta$ with a difference of ten times order of magnitude. It is seen that as the value of $\beta$ becomes smaller, the results of GUP
more coincide with the results of the commutative case. Considering a very small value for the parameter $\beta$, for example suggested by string theory, reveals that at least at the present age of the universe ($t\simeq \omega^{-1}$) it is impossible to distinguish between the GUP and commutative cases in that which approach fits better the data.

\section{Quantum solutions}
\subsection{Commutative case}

To investigate the quantum version of the model, we use the WD
equation, ${\cal H}\Psi=0$. Here ${\cal H}$ is the operator form
of the Hamiltonian (\ref{e13}). Using the canonical procedure to
quantum mechanics that the phase space variables are replaced with
quantum operators, by replacing $p_i=-i\frac{\partial}{\partial
x_i}$ in (\ref{e13}) we get the following WD equation:
\bea\label{e43} \Bigg[-\frac{\partial^2}{\partial
x_1^2}+\frac{\partial^2}{\partial
x_2^2}+4\omega^2(x_1^2-x_2^2)\Bigg]\Psi(x_1,x_2)=0.
 \eea
This equation is the quantum version of oscillator-ghost
oscillator system with zero energy condition. We can solve it by
Formal variable separation approach as:
 \bea\label{e44}
\Psi_{n_1,n_2}(x_1,x_2)=U_{n_1}(x_1)V_{n_2}(x_2),
 \eea

 \bea\label{e45}
 \frac{\partial^2W_i}{\partial
x_i^2}+(\lambda-4\omega^2x_i^2)W_i=0,&        & W_i(i=1,2)=U,V
  \eea
in negative cosmological background, the solutions are
 \bea\label{e46}
U_{n_1}(x_1)=\Bigg(\frac{2\omega}{\pi}\Bigg)^{1/4}\frac{e^{-\omega
x_1^2}}{\sqrt{2^{n_1}n_1!}}H_{n_1}(\sqrt{2\omega}x_1),
 \eea
 \bea\label{e47}
V_{n_2}(x_2)=\Bigg(\frac{2\omega}{\pi}\Bigg)^{1/4}\frac{e^{-\omega
x_2^2}}{\sqrt{2^{n_2}n_2!}}H_{n_2}(\sqrt{2\omega}x_2),
 \eea
with $\lambda=2n_i+1$ and the restriction $n_1=n_2=n$. Here,
$H_n(x)$ are Hermite polynomial and the eigenfunction are
normalized as:
 \bea\label{e48}
 \int_{-\infty}^{+\infty}e^{-x^2}H_n(x)H_m(x)dx=2_n\pi^{1/2}n!\delta_{nm}.
 \eea
The general solution of the WD equation can be written as a
superposition of the above eigenfunctions:
\bea\label{e49}
 \Psi(x_1,x_2)=\Big(\frac{2\omega}{\pi}\Big)^{1/2}e^{-\omega(x_1^2+x_2^2)}\sum_{n}\frac{c_n}{2^nn!}
 H_n(\sqrt{2\omega}x_1)H_n(\sqrt{2\omega}x_2).
 \eea
Coefficients $c_n$ are so chosen as to make the states coherent
\cite{Gousheh}:
\bea\label{e50}
 c_n=\Bigg(\frac{\pi}{2\omega}\Bigg)^{\frac{1}{4}}\frac{\chi_0^n \frac{n}{2}!}{{(-1)}^{\frac{n}{2}}n!}e^{-\frac{1}{4}|\chi_0|^2}.
 \eea

\begin{figure}
\vskip 1.5 cm
    \includegraphics{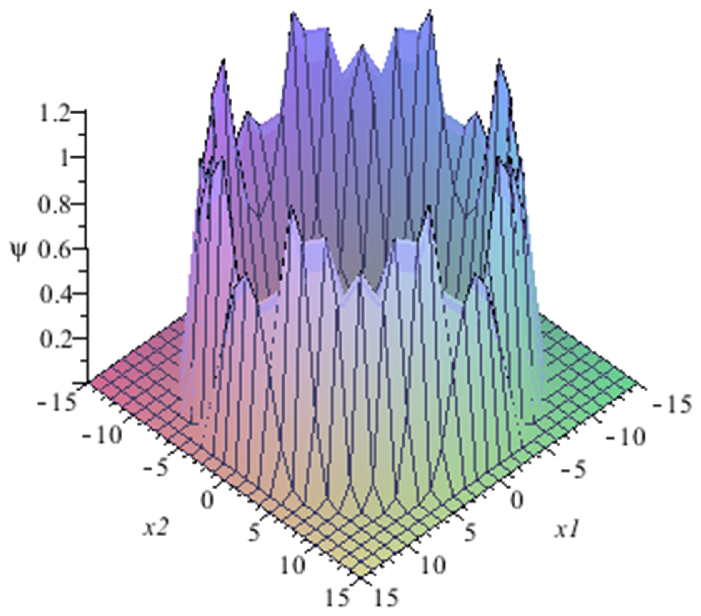}
    \includegraphics{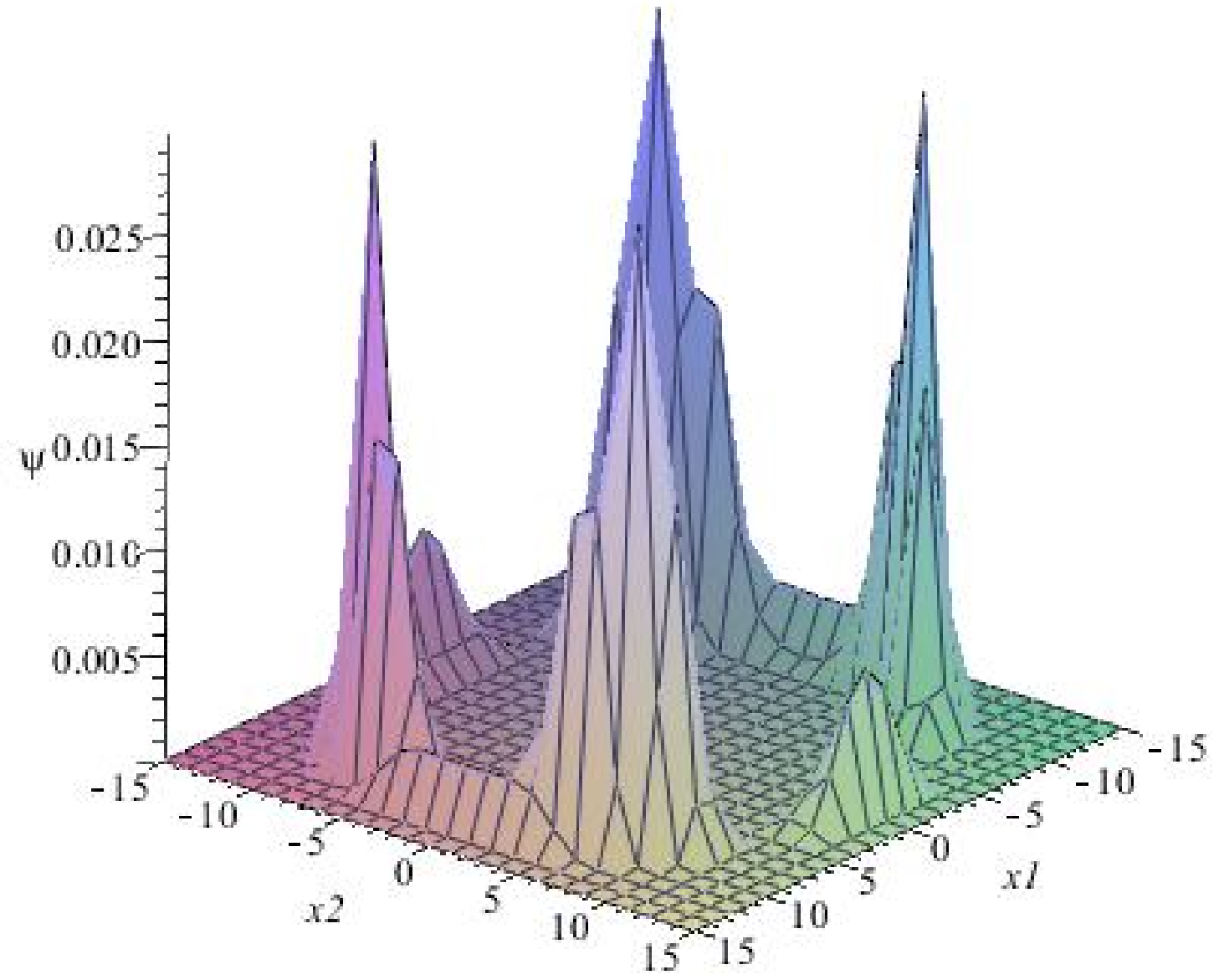}
\vskip 2.5cm \caption{ The square of the wavefunction of the
universe in $ x_1-x_2 $ plane with one extra dimension and
negative cosmological constant. Left and right figures refer to
the commutative and GUP framework, respectively. }
\end{figure}

In general, the quantum solutions do not offer semiclassical
description of some space-time region unless we introduce a
decoherence mechanism which is widely regarded as necessary to
assign a probability for the occurrence of a classical space-time.
However, in the lack of desired decoherence mechanism, in order
for a satisfactory classical-quantum correspondence is achieved we
may at least investigate if the absolute values of the solutions
of Wheeler-DeWitt equation have maxima in the vicinity of the
classical loci. Fig.7 (left) shows the square of the wavefunction
in the case of negative cosmological constant. As we can see from
this figure and the corresponding classical trajectory in Fig.6
(left), the peak follows the classical trajectory. Thus, it is
seen that there is a good correspondence between quantum
description and the classical pattern of the model.

We get the corresponding answers in the positive cosmological
background by replacing $\omega^2$ with $-\omega^2$ in the above
equations. Fig.7, shows the square of the wavefunction. Because of
highly increasing dependence of the amplitude of the wavefunction
on $x_1$ and $x_2$, the fluctuation of wavefunction on the line
$x_1=x_2$ can not be seen using the normal scales adopted in this
figure. However, the peak of squared wavefunction follows properly
the classical trajectory, namely the line $x_1=x_2$ in Fig.6
(right). Therefore, similar to the case of negative cosmological
constant, there is a good correspondence between quantum
description and the classical pattern.

\subsection{GUP case}

Here, we aim to investigate influence of GUP in quantum
cosmological model that has been presented above. The Hamiltonian
of the model is given by (\ref{e13}). To construct WD equation in
GUP framework we use Eq.(\ref{e32}) with the representation
$p_{i0}=-i \frac{\partial}{\partial x_{i0}}$ for the momentum
operator. In first order in $\beta$, by ignoring zero subscript we
have

\bea\label{e51} \Bigg[2\beta \Bigg(\frac{\partial^4}{\partial
x_1^4}+ \frac{\partial^4}{\partial
x_2^4}-2\frac{\partial^2}{\partial
x_1^2}\frac{\partial^2}{\partial
x_2^2}\Bigg)+\left(-\frac{\partial^2}{\partial
x_1^2}+\frac{\partial^2}{\partial
x_2^2}\right)+4\omega^2(x_1^2-x_2^2)\Bigg]\Psi(x_1,x_2)=0.
 \eea

We see that this is not a separable differential equation, so we set
a procedure to convert it into a separable one. Up to first order in
$\beta$, we assume a perturbative solution, $\Psi=\Psi^{(0)}+\beta
\Psi^{(1)}$, where $\Psi^{(0)}$ is the wave function of
commutative case, namely the solution of Eq.(\ref{e43}). Substitution of
this perturbative solution into Eq.(\ref{e51}) reveals that due to the presence of $\beta$ in the first order part of Eq.(\ref{e51}) we may put $\Psi\simeq\Psi^{(0)}$, hence we can use the commutative case equation (\ref{e43}) for this part. By this consideration we have

\bea\label{e51-2} \Bigg[2\beta \Bigg(\frac{\partial^4}{\partial
x_1^4}+ \frac{\partial^4}{\partial
x_2^4}&-&2\frac{\partial^2}{\partial
x_2^2}\left(\frac{\partial^2}{\partial
x_2^2}+4\omega^2(x_1^2-x_2^2)\right)\Bigg)
\nnb \\
&+&\left(-\frac{\partial^2}{\partial x_1^2}+\frac{\partial^2}{\partial
x_2^2}\right)+4\omega^2(x_1^2-x_2^2)\Bigg]\Psi(x_1,x_2)=0,
 \eea

\bea\label{e51-4} \Bigg[2\beta \Bigg(\frac{\partial^4}{\partial
x_1^4}- \frac{\partial^4}{\partial
x_2^4}&-&8\omega^2x_1^2\frac{\partial^2}{\partial
x_2^2}+8\omega^2\frac{\partial^2}{\partial x_2^2}(x_2^2)\Bigg)
\nnb \\
&+&\left(-\frac{\partial^2}{\partial x_1^2}+\frac{\partial^2}{\partial
x_2^2}\right)+4\omega^2(x_1^2-x_2^2)\Bigg]\Psi(x_1,x_2)=0.
 \eea

Because of the term $x_1^2\frac{\partial^2}{\partial x_2^2}$, the
equation is inseparable too. To resolve the problem we use the
classical loci $x_1^2+x_2^2=r^2$ (see figure 6).

\bea\label{e51-4'} \Bigg[2\beta \Bigg(\frac{\partial^4}{\partial
x_1^4}- \frac{\partial^4}{\partial
x_2^4}-8\omega^2(r^2-x_2^2)\frac{\partial^2}{\partial
x_2^2}+8\omega^2\frac{\partial^2}{\partial
x_2^2}(x_2^2)\Bigg)+\left(-\frac{\partial^2}{\partial
x_1^2}+\frac{\partial^2}{\partial
x_2^2}\right)+4\omega^2(x_1^2-x_2^2)\Bigg]\Psi=0.\nnb
 \eea

Separation of the variables as $\Psi(x_1,x_2)=U(x_1)V(x_2)$ with
perturbative solutions in first order of $\beta$ as
$U(x_1)=U^{(0)}(x_1)+\beta U^{(0)}(x_1)$ and
$V(x_2)=V^{(0)}(x_2)+\beta V^{(0)}(x_2)$, and perturbative separation of
the separation constant $\lambda=\lambda^{(0)}+\beta \lambda^{(1)}$ leads to

\begin{itemize}
 \item in zeroth order in $\beta$
\bea\label{e53}
 && \frac{\partial^2W^{(0)}(x_i)}{\partial
x_i^2}+(\lambda^{(0)} -4\omega^2x_i^2)W^{(0)}(x_i)=0,
\\ \nonumber
&& W^{(0)}(x_i;i=1,2)=U^{(0)}(x_1),V^{(0)}(x_2),
  \eea

 \item in first order in $\beta$.

 \bea\label{e54}
&& \frac{\partial^2W^{(1)}(x_i)}{\partial
x_i^2}+(\lambda^{(0)}-4\omega^2x_i^2)W^{(1)}(x_i)=g(x_i),
\\ \nonumber
&& W^{(1)}(x_i;i=1,2)=U^{(1)}(x_1),V^{(1)}(x_2),
\\ \nonumber
&& g(x_1)=2 \frac{\partial^4U^{(0)}(x_1)}{\partial
x_1^4}-\lambda^{(1)}U^{(0)}(x_1),
\\ \nnb
&& g(x_2)=2\Bigg(\frac{\partial^4V^{(0)}}{\partial
x_2^4}+8\omega^2(r^2-x_2^2)\frac{\partial^2V^{(0)}}{\partial
x_2^2}-8\omega^2\frac{\partial^2(x_2^2 V^{(0)})}{\partial
x_2^2}\Bigg)-\lambda^{(1)}V^{(0)}.
  \eea

\end{itemize}

\begin{figure}
\vskip 1.5 cm
    \includegraphics{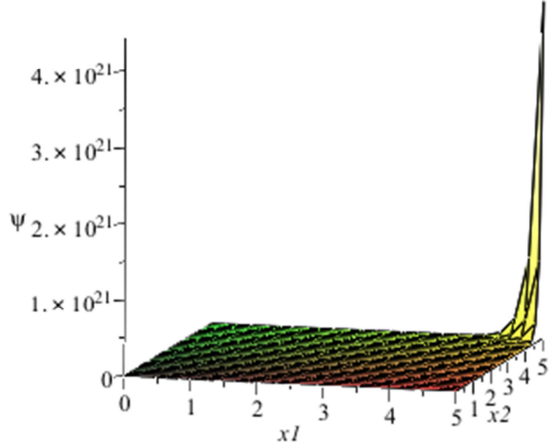}
\vskip 2.5cm \caption{The square of the wavefunction of universe
in $ x_1-x_2 $ plane with one extra dimension and positive
cosmological constant.}
\end{figure}

We can see that Eq.(\ref{e53}) is commutative limit
$(\beta\rightarrow 0)$ of Eq.(\ref{e51}) (see Eq.(\ref{e45})). So
it's solution is

\bea\label{e55}
W^{(0)}_{n}(x_i)=\Bigg(\frac{2\omega}{\pi}\Bigg)^{1/4}\frac{e^{-\omega
x_i^2}}{\sqrt{2^{n}n!}}H_{n}(\sqrt{2\omega}x_i).
 \eea
Eq.(\ref{e54}) is a inhomogeneous differential equation. The solution of homogeneous part is

 \bea\label{e56}
W^{(1)}_{n}(H)&=&\Bigg(\frac{2\omega}{\pi}\Bigg)^{1/4}\frac{e^{-\omega
x_i^2}}{\sqrt{2^{n}n!}}H_{n}(\sqrt{2\omega}x_i)
\nnb \\
     &=&W^{(0)}_{n}(x_i).
 \eea
By use of Eq.(\ref{e56}), the solution of inhomogeneous part can be written as

 \bea\label{e57}
W^{(1)}_{n}(I)=\xi_1(x_i)W_0^{(0)}(x_i)+\xi_2(x_i)W_n^{(0)}(x_i),
\eea
where
\bea\label{e58}
\xi_1(x_i)&=&\frac{1}{\mu(x_i)}\int^{x_i}\mu(x_i)\frac{g_nW_n^{(0)}}{W^{(0)'}_0W^{(0)}_n-W^{(0)'}_nW^{(0)}_0} dx_i,
\nnb \\
\xi_2(x_i)&=&\int^{x_i}\frac{2nW_0^{(0)2}\xi_1-g_nW_0^{(0)}}{W^{(0)'}_0W^{(0)}_n-W^{(0)'}_nW^{(0)}_0}dx_i,
\nnb \\
\mu(x_i)&=&Exp\Bigg(\int^{x_i}\frac{2nW_0^{(0)}W_n^{(0)}}{W^{(0)'}_0W^{(0)}_n-W^{(0)'}_nW^{(0)}_0}dx_i\Bigg),
\eea
and a prime denotes for differentiation with respect to $x_i$.
Combination of zeroth and first order solutions leads to

 \bea\label{e59}
 W_n(x_i)=W_n^{(0)}(x_i)+\beta(W_n^{(0)}(x_i)+\xi_1(x_i)W_0^{(0)}(x_i)+\xi_2(x_i)W_n^{(0)}(x_i)).
\eea
Therefore, we may write the general solution of WD equation in the GUP framework as a superposition of above eigenfunctions

\bea\label{e60}
 \Psi_{GUP}(x_1,x_2)=\sum_{n}c_n W_n(x_1)W_n(x_2).
 \eea
We set, as in Eq.(\ref{e50}), the coefficient $c_n$ to the coefficient of commutative wavefunction. Figure.7 (right) shows the square of the wavefunction with negative cosmological constant in GUP framework. As is seen in this figure, there are four peaks distributed around $x_1, x_2=\pm8.5$ which in the classical viewpoint correspond to a state with $R_{max}$. Also, the overall peak follows the classical trajectory in figure.6 (left). Thus, similar to
the commutative case, there is a good correspondence between quantum description and the classical pattern of the model in GUP case.

\section{Problem of time}

The canonical approach at the classical level gives the Hamiltonian constraint. Promoting this constraint to the quantum level does not give a time-dependent
wave equation, rather leads to a stationary and timeless equation so called
Wheeler-DeWitt equation. This suggests that noting at all evolves in the universe. Therefore, one is faced with the problem of explaining the origin of the notions of time in the laws of physics. This is so called time problem in quantum gravity and there are different approaches to solve this problem
\cite{Pt}. Among these approaches, a new and relevant approach to this paper is the one introduced in \cite{khosravi2}-\cite{khosravi3} where the authors offer an explanation of time evolution in quantum gravity by using of the probability density in quantum mechanics. They establish a method called
"Probabilistic Evolutionary Process" (PEP) which states that if the system is in a given quantum state, then a small perturbation inspires this
state to end up in another quantum state with the higher probability of existence. The evolutionary process of initial quantum states toward the more probable final quantum states suggests a possible solution to the problem of time in quantum cosmology. 

In our problem by consideration of time
evolution in classical context, which accords with increasing in
$x_1$ and $x_2$, we may say that such a transition in quantum
state from low probability to high probability, can be considered
as a quantum evolutionary process. This interpretation of time
evolution may be justified in the positive cosmological constant
case, where there is considerable difference in the probability
amplitude of wavefunction. However, because of nearly equal
probability amplitude of wavefunction in the negative cosmological
constant case, this interpretation of time evolution is not of
practical use to solve the time problem.

In classical context, the time evolution is in accordance
with going from $R_{min}$ to $R_{max}$. As we mentioned in previous
subsection, by consideration of higher probability of existence
in quantum mechanics and using the discussion in references \cite{khosravi2}-\cite{khosravi3},
we may explain the time problem of the WD equation in quantum gravity.
It should be noted that, this interpretation is due to the influence
of GUP.

\section{Conclusion}
We have considered a multidimensional cosmology having FRW type
metric of 4-dimensional space-time and $d$-dimensional Ricci-flat
internal space coupled with a higher dimensional cosmological
constant. The classical cosmology in commutative and GUP cases are
studied in detail and the corresponding exact solutions for
negative and positive cosmological constants are obtained. For
negative cosmological constant, both cases result in finite size
universes which differ by their size and age, while, for positive
cosmological constant both cases result in infinite size universes
having late time accelerating behavior in good agreement with the
current observations. Both commutative and GUP cases with negative
and positive cosmological constants result in the stabilization of
internal space to the sub-Planck size. We have also derived the
Wheeler-DeWitt equation and obtained the solutions in both cases
for both negative and positive cosmological constants. It is shown
that good correspondence exists between the classical and quantum
solutions.

It is worth noting that the effects of GUP in this model are
important not only in the early universe but also in the late time
behavior of the cosmic evolution. So, in the GUP framework, we can
see somehow the indirect quantum gravitational effects at large
scale. Moreover, the influence of GUP to solve the time problem in
this model is remarkable.

\section{Appendix}

We want to show that the momentum $p_j$ in the GUP framework in
terms of the ordinary low energy momentum $p_{j0}$, equation
(\ref{e31.2}) fulfill the commutation relations (\ref{e29}) and
(\ref{e30}) with $\beta'=2\beta$ up to first order in $\beta$. Let
us compute (\ref{e29})

\bea\label{e61}
 [x_i, p_j]&=&[x_{i0}, p_{j0}+\beta p_{j0}p_{k0}p_{k0}]=[x_{i0}, p_{j0}]+\beta[x_{i0},  p_{j0}p_{k0}p_{k0}]
  \nnb \\
 &=&i \delta_{ij}+\beta p_{j0}[x_{i0},  p_{k0}p_{k0}]+\beta[x_{i0},  p_{j0}]p_{k0}p_{k0}
  \nnb \\
 &=&i \delta_{ij}+2i \beta \delta_{ik}p_{k0}p_{j0}+i \beta\delta_{ij}p_{k0}p_{k0}
 \nnb \\
 &=&i (\delta_{ij}+\beta \delta_{ij}p_0^2+2 \beta p_{i0}p_{j0}),
 \eea
 where we have used the identity $[P, QS]=Q[P, S]+[P, Q]S$. Up to
 first order in $\beta$ it is obvious that

\bea\label{e62} \beta p^2=\beta p_i p_i=\beta p_{i0}+\beta
p_{i0}p_{k0}p_{k0}=\beta p_{i0} p_{i0}=\beta p_0^2,
\eea
and similarly $\beta p_{i} p_{j}=\beta p_{i} p_{j}$. Substituting
these in equation (\ref{e61}) yields

\bea\label{e63}
[x_i, p_j]=i (\delta_{ij}+\beta \delta_{ij}p^2+2
\beta p_{i}p_{j}),
 \eea
which is nothing but Eq.(\ref{e29}) with $\beta'=2\beta$. Eq.
(\ref{e30}) is a direct consequence of $[p_{i0}, p_{j0}]=0$.


\newpage
\begin{thebibliography}{99}

\bibitem{string1} G. Veneziano, Europhys. Lett. 2 (1986) 199; Proc. of Texas Superstring Workshop (1989); D. Gross, Proc. of ICHEP, Munich (1988).

\bibitem{string2}D. Amati, M. Ciafaloni and G. Veneziano, Phys. Lett. B216 (1989) 41.

\bibitem{string3}D. Amati, M. Ciafaloni and G. Veneziano, Phys. Lett. B197 (1987) 81; Int. J. Mod. Phys. A3 (1988) 1615; Nucl. Phys. B347 (1990) 550.

\bibitem{string4}D. J. Gross and P. F. Mende, Phys. Lett. B197 (1987) 129;
Nucl. Phys. B303 (1988) 407.

\bibitem{string5} M. Ciafaloni, 'Planckian Scattering beyond the Eikonal Approximation', preprint DFF 172/9/'92 (1992).
International Workshop on Theoretical Physics "Ettore Majorana", Erice, Italy, 21 - 28 Jun 1992, pp.249.

\bibitem{string6}K. Konishi, G. Paffuti and P. Provero, Phys. Lett. B234 (1990) 276.

\bibitem{general GUP1} A. Kempf, G. Mangano and R. B. Mann, Phys. Rev. D52 (1995) 1108. 

\bibitem{sepangi} H. R. Sepangi, B. Shakerin, B. Vakili, Class. Quant. Grav.
26 (2009) 065003.

\bibitem{general GUP2} A. Kempf and G. Mangano, Phys. Rev. D55 (1997) 7909. 

\bibitem{general GUP3} M. Maggiore, Phys. Lett. B319 (1993) 83;
 M. Maggiore, Phys. Lett. B304 (1993) 65.

\bibitem{general GUP4} A. Kempf, J. Math. Phys.35 (1994) 4483;
A. Kempf and J. C. Niemeyer, Phys. Rev. D64 (2001);
L. N. Chang, D. Minic, N. Okamura and T. Takeuchi, Phys. Rev. D65 (2002) 125027.

\bibitem{general GUP5} R. J. Adler, D. I. Santiago, Mod. Phys. Lett. A14 (1999) 1371.

\bibitem{general GUP6}F. Scardigli, Phys. Lett. B452 (1999) 39.

\bibitem{general GUP7}S. F. Hassan, M. S. Sloth, Nucl. Phys. B674 (2003) 434.

\bibitem{general GUP8}A. Kempf, L. Lorenz, Phys. Rev. D74 (2006) 103517.

\bibitem{Snyder} H. Snyder, Phys. Rev. 71 (1947) 38.

\bibitem{recent past} A. Connes, Noncommutative Geometry, Academic, New York, 1994; A. Connes J. Math. Phys. (N.Y.) 41 (2000) 3832;
J. C. Varilly, An Introduction to Noncommutative Geometry [arXiv: physics/9709045];
M. R. Douglas and N. A. Nekrasov, Rev. Mod. Phys. 73 (2001) 977.

\bibitem{Duff} M. J. Duff, B. E. W. Nilsson, and C. N. Pope,Kaluza-Klein supergravity, Phys. Rep.130, 1-142 (1986).

\bibitem{Schwarz} J. H. Schwarz, Superstrings (World Scientific, Singapore, 1985).

\bibitem{Kaluza} T. Kaluza, Zum Unitatsproblem der Physik, Sitz. Preuss. Akad. Wiss. Phys. Math. K1, (1921) 966.

\bibitem{Klein}O. Klein, Quantentheorie und funfdimensionale Relativitatstheorie, Zeits. Phys. 37, (1926) 895.

\bibitem{Ext1}M. Cavaglia', S. Das, R. Maartens, Class. Quant. Grav.20 (2003) L205.

\bibitem{Ext2}F. Scardigli, {\it Glimpses on the micro black hole Planck phase}, (arXiv:0809.1832).

\bibitem{khosravi1} N. Khosravi,S. Jalalzadeh, H. R. Sepangi, JHEP 01, (2006)
134.

\bibitem{maxa1} T. Kakuda, K. Nishiwaki, K-y. Oda, N. Okuda, R. Watanabe;
 Proceedings of International Linear Collider Workshop (LCWS11), 26-30 September 2011, Granada, Spain [arXiv:hep-ph/1202.6231v1]

\bibitem{Pt}C. J. Isham, {\it Canonical Quantum Gravity and the Problem of
Time}, gr-qc/9210011; J. J. Halliwell, {The Interpretation of Quantum Cosmology and the Problem of Time}, gr-qc/0208018; E. Anderson, {\it The Problem of Time in Quantum Gravity}, arXiv:1009.2157.

\bibitem{khosravi2} N. Khosravi,S. Jalalzadeh, H. R. Sepangi, Gen. Relativ. Gravit.39 (2007) 899.

\bibitem{khosravi3}N. Khosravi, H. R. Sepangi, Phys. Lett. B673 (2009) 297.


\bibitem{Matej Pavsic} M. Pavsic, Phys. Lett. A254 (1999) 119.


\bibitem{Majumder} Barun Majumder, Phys. Rev. D84, 064031 (2011).


\bibitem{S. Benczik}S. Benczik, et al., Phys. Rev. D66 (2002) 026003; S. Benczik, et al., Classical Implications of the Minimal Length Uncertainty Relation [arXiv: hep-th/0209119].

\bibitem{Lay}L. N. Chang, D. Minic, N. Okamura, and T. Takeuchi, Phys. Rev. D65 (2002) 125028.

\bibitem{Gousheh} S. S. Gousheh,  H. R. Sepangi, Phys. Lett. A272, (2000)
304.

\end{thebibliography}
\end{document}